# Irreversible Temperature Quenching and Antiquenching of Photoluminescence of ZnS/CdS:Mn/ZnS Quantum Well Quantum Dots

X. Ding[1,a)], R.C. Dai[2,a)], Z. Zhao[3], Z.P. Wang[2], Z.Q. Sun[1], Z.M. Zhang[2b)] and Z.J. Ding[3c)]

[1] School of Physics and Material Science, Anhui University, Hefei, Anhui 230601, P. R. China

[2] Centre of Physical Experiments, University of Science and Technology of China, Hefei, Anhui 230026, P. R. China

[3] Hefei National Laboratory for Physical Sciences at Microscale and Physics Department, University of Science and Technology of China, Hefei, Anhui 230026, P. R. China

An experimental observation on irreversible thermal quenching and antiquenching behavior is reported for photoluminescence (PL) of ZnS/CdS:Mn/ZnS quantum well quantum dots (QWQDs) prepared with a reverse micelle method. The dual-color emissions, a blue emission band centered at 430 nm and a $Mn^{2+}$ $^4T_1 \rightarrow {}^6A_1$ orange emission peak at 600 nm, were found to have different dependences of emission intensity on temperature in the range of 8-290 K. Depending on $Mn^{2+}$ doping concentration, they can both show strong antiquenching behavior in a certain temperature range in addition to the usual thermal quenching when lowering down temperature while it is very prominent for orange emission but weaker for blue emission. Moreover, the antiquenching behavior is weakened with rise of temperature, giving rise to a hysteretic PL temperature dependence for the QWQDs.

a) X. Ding and R.C. Dai contributed equally to this work
b) email: zzm@ustc.edu.cn
c) email: zjding@ustc.edu.cn

Colloidal semiconductor nanocrystals with size tunable bandgap have attracted great research interest for the technical applications to photovoltaic cells,[1,2] light-emitting diodes (LEDs),[3-6] lasers[7,8] and biological labels.[9] The II-VI semiconductors quantum dots (QDs) exhibit excellent optical properties for photoluminescence (PL); by the quantum size effect the band gap can be tailored by controlling crystal size, leading to tunable emission and excitation wavelengths.[10] It has been known that doping with atomic impurities is an effective way to manipulate luminescence properties.[11] $Mn^{2+}$ doped ZnS QDs show a single characteristic orange emission peak at 590-600 nm related to $^4T_1 \rightarrow {}^6A_1$ transition with quantum yield (QY) of 18%.[12] In addition, ZnS:Mn QDs may have an additional blue emission centered at 420-430 nm whose bandwidth is much larger than orange emission.[13-15] The dual-color property has been also attained for ZnS:Mn quantum rods.[16]

On the other hand, because of large surface-to-volume ratio for QDs the surface-related trap states may act as fast non-radiative de-excitation channels for photo excited charge carriers to be the luminescence quenching centers. The luminescent properties can then be dramatically improved by epitaxially growing an inorganic shell of a wider band gap semiconductor on the core surface. Such a shell provides a potential barrier to confine charge carriers in the core region so as to passivate surface-related defect states, leading to more efficient and photostable luminescence QDs[10]. A bright blue luminescence of 460-480 nm with QY of 20-30% has obtained for CdS/ZnS core-shell QDs.[17] For Mn doped ZnS core-shell QDs, the blue emission and the orange $Mn^{2+}$ emission can be simultaneously enhanced by passivation of ZnS shell with the same bandgap as ZnS core.[18,19] Therefore, CdS:Mn/ZnS QDs are shown to have better improved luminescent properties for $Mn^{2+}$ emission[20-22] by providing a larger bandgap of ZnS shell than CdS core. Furthermore, core-shell-shell heteronanostructures are developed[10] and ZnS/CdS/ZnS quantum well quantum dots (QWQDs) with a large lattice mismatch have been successfully fabricated.[23] The $Mn^{2+}$ doped QWQDs, ZnS/CdS:Mn/ZnS, are shown to have even a brighter orange emission, with a weak blue band at 460-480 nm at room temperature, than the CdS:Mn/ZnS QDS and the QY of 45% was attained.[24]

The temperature dependence of PL is often used to obtain information about electronic gap levels in semiconductors, the radiative and nonradiative relaxation processes. Usually the integrated PL intensity is reduced as temperature increases because of increasing transition rate via nonradiative processes, as for emissions from several different core-shell QDs[25-27] and for $Mn^{2+}$ emission of CdS:Mn QDs.[28] The thermal quenching in these QDs is resulted from carrier trapping by surface defect states and thermal escape assisted by the scattering with multiple LO phonons. Recently an anomalous thermal quenching behavior, i.e. antiquenching, has been found on the orange emission of CdSe QDs which are capped different ligand shells. The thermal antiquenching temperature is within a range of 20-50 K below room temperature and the intensity variation is around 10-35%, depending on the number of carbon atoms in the capping aliphatic molecules;[29] the mechanism is considered to be related with a phase transition of ligand shell. Nevertheless, the antiquenching has also been found the orange emission from CdSe/CdS core-shell and CdSe/CdS/CdZnS/ZnS core-multishell QDs where the origin of antiquenching was attributed to the thermal activation of carriers from trapping states localized at interfaces.[30] In this letter, we report an irreversible thermal quenching and antiquenching behavior for blue and orange emissions from ZnS/CdS:Mn/ZnS QWQDs.

ZnS/CdS:Mn/ZnS QWQDs were synthesized via a reverse micelle method.[24] In this process, standard solutions of $Zn^{2+}$ (0.26 M), $Cd^{2+}$ (0.10 M), $Mn^{2+}$ (0.10 M) and $S^{2-}$ (2.30 M) were prepared from $Zn(CH_3COO)_2 \cdot 2H_2O$, $Cd(CH_3COO)_2 \cdot 2H_2O$, $Mn(CH_3COO)_2 \cdot 2H_2O$ and $Na_2S$. AOT (dioctylsulfosuccinate sodium salt) was dissolved in heptane to prepare 0.10 M AOT/heptane as a surfactant solution. The appropriate volumes of aqueous $Zn(CH_3COO)_2$ solution was mixed with the AOT/heptane stock solution to produce water/oil microemulsions. The molar ratio of water to surfactant is 10 for all micellar solutions. The ($Cd^{2+}+Mn^{2+}$) and $S^{2-}$ microemulsions were prepared in the same way. The $Mn^{2+}/Cd^{2+}$ molar ratio, $x$%, was obtained by adding appropriate standard solutions of $Mn^{2+}$(0.10 M) into $Cd^{2+}$(0.10 M) solutions. First, ZnS/CdS:Mn core-shell QDs are formed by mixing $S^{2-}$ and $Zn^{2+}$ containing micellar solution of 100 ml for 15 min. For growth of CdS:Mn shell onto the ZnS core, 120 ml of ($Cd^{2+}+Mn^{2+}$) containing micellar solution was added into the ZnS core micellar solution at a slow rate of ~1 ml/min. Then, 250 ml of $Zn^{2+}$ containing micellar solution was injected into ZnS/CdS:Mn micellar solutions and mixed for 8 h at room temperature (10-20 °C) to form the ZnS/CdS:Mn/ZnS QWQDs. After thorough washing with heptane and methanol, the QWQDs were dispersed in methanol. By varying $Mn^{2+}/Cd^{2+}$ molar ratio, a series of samples containing different initial $Mn^{2+}$ concentrations (mol%) of 2%, 4%, 6%, 8% and 10% are obtained. Sample of $x\%$ $Mn^{2+}$ concentration is referred as $S_{x\%}$.

The x-ray diffraction (XRD) pattern of the synthesized sample was obtained with an x-ray diffraction facility (TTR-III). UV-Vis absorption spectra were recorded by a Shimadzu DUV-3700 spectrophotometer. The photoluminescence excitation (PLE) and PL spectra were taken on a steady-state/lifetime spectrofluorimeter (FLUOROLOG-3-TAU, Jobin Yvon) with a 450 W xenon lamp as the excitation source. Temperature dependent PL measurements were performed with the spectrofluorimeter in the temperature range of 8-290 K. XRD pattern in the inset of Fig. 1(a) shows three diffraction peaks which are indexed to be cubic zinc blende structure as in the standard card (JCPDS No.77-2100). Positions of diffraction peaks of the QWQDs are close to the bulk ZnS and the peak broadening is due to nanocrystalline nature of the sample. Fig. 1(a) shows the absorbance spectra of as-grown ZnS/CdS:Mn/ZnS QWQDs. All the samples have similar spectrum feature, and the absorbance edges are within 330-370 nm (3.76-3.35 eV). The onsets of the spectra are quite close to that of 2.6 nm ZnS/CdS/ZnS QWQDs[22] showing the similar size of nanoparticles; while the variation of onsets indicates that there is a certain fluctuation on size distribution between samples of different $Mn^{2+}$ concentrations. Although the inhomogeneous size distribution of QWQDs in a sample can cause a broadening of the onset position, however, the tight-binding calculation for 2.6 nm ZnS/CdS/ZnS QWQDs presents the transition energy about 3.39-3.77 eV,[22] which is also a good indication of the size range of 2-3 nm for the prepared QWQDs.

Fig. 1(b) presents the room temperature PLE and PL spectra of ZnS/CdS:Mn/ZnS QWQDs. The PLE spectra were obtained by monitoring the orange emission of $Mn^{2+}$ at 600 nm. The PLE spectrum consists of a strong absorption peaks at 340 nm and one shoulder at 400 nm for all the samples except $S_{10\%}$ for which the shoulder is quite broadened. Two peaks at 340 nm and 400 nm are associated with band absorption from ZnS core and CdS inner shell, respectively.[24] The relative intensity of the 340 nm peak is much stronger than the 400 nm in the PLE spectrum, indicating that energy transfer is dominantly from ZnS to $Mn^{2+}$ in ZnS/CdS:Mn/ZnS QWQDs. The 340 nm peak intensity rises with $Mn^{2+}$ concentration. The PL spectra were recorded at the excitation wavelength of 325 nm. The 600 nm peak intensity of the QWQDs increases with $Mn^{2+}$ concentration to a maximum for $S_{8\%}$; combing with PLE spectra this feature indicates that the $Mn^{2+}$ doping level in QWQDs is proportional to initial $Mn^{2+}$ concentration. Meanwhile the peak position is slightly red shifted with increasing $Mn^{2+}$ doping. In the case of ZnS:Mn nanocrystals the 600 nm orange luminescence originates from $Mn^{2+}$ ions on $Zn^{2+}$ sites, where the $Mn^{2+}$ is tetrahedrally coordinated by $S^{2-}$.[13] Mechanistic study on the Mn-doping of CdS/ZnS core-shell QDs by a three-step synthesis method has shown that, $Mn^{2+}$ ions are firstly adsorbed on the CdS-core surface to form weakly bound $Mn^{2+}$ and then followed by a chemical reaction to form the strongly bound $Mn^{2+}$; both weakly bound and strongly $Mn^{2+}$ can be removed during ZnS-shell growth, while the replacement of weakly bound $Mn^{2+}$ can occur below 240 °C while it requires higher temperature for strongly bound $Mn^{2+}$.[31] It is then highly possible that, for our QWQDs synthesized at a lower room temperature (10-20 °C) considerable strongly bound $Mn^{2+}$ ions are located at ZnS/CdS inner interface with some $Mn^{2+}$ ions replaced the $Cd^{2+}$ sites in the CdS middle shell.

Fig. 2 shows the temperature dependence of PL spectra of ZnS/CdS:Mn/ZnS QWQDs for samples, $S_{2\%}$, $S_{4\%}$ and $S_{8\%}$, in the temperature cooling down process and heating up process in temperature range of 8-290 K. At each temperature a relaxation time of 10 min was held before the PL spectrum measurement. For sample $S_{2\%}$ there is only orange emission of $Mn^{2+}$ and its temperature dependence is basically the thermal quenching behavior although a small hysteresis exists. But for other samples having additional blue emission the PL temperature behavior is anomalous and quite complex: First, the blue emission has rather different temperature varying tendency with the orange emission; Second, in a certain temperature region the emission intensity can be decreased with reducing temperature, presenting an antiquenching effect. Finally, an obvious hysteresis of PL spectrum intensity has been found, which is particularly strong for the orange emission. For example, for $S_{4\%}$, the room temperature emission intensity is largely lost after returning temperature from low temperature. This is an unusual irreversible phenomenon found on the ZnS/CdS:Mn/ZnS QWQDs. The $Mn^{2+}$ emission is red shifted and the blue emission is blue shifted with lowering temperature as clearly seen in Fig. 2(c), in agreement with previous observation for ZnS:Mn QDs.[14]

The emission peak intensities have then been integrated to show clearer temperature dependence. Both orange and blue emission peak were fitted with two Gaussian terms. Fig. 3 displays the integrated PL intensities as functions of temperature for all the samples measured. Evident quenching and antiquenching are found in the cooling and heating processes; the largest intensity variation in antiquenching can even reach 50%. But it is amazing that the transition from quenching behavior to antiquenching behavior is not in the sequence of $Mn^{2+}$ doping level. For $S_{6\%}$, the 430 nm blue emission is quenching behavior while the orange emission shows antiquenching down to 50 K; the orange emission is quasi-reversible and is even over-recovered at the room temperature. For $S_{4\%}$, the blue emission is firstly quenching in the cooling process from 290 K down to 175K and has very strong antiquenching in range of 35-175 K; in heating process from 8 K it is firstly quenched to 50 K and is then followed by an antiquenching in 50-175 K and quenching in 175-290 K. There are two transitions between antiquenching and quenching. For the orange emission the antiquenching is found in very low temperature region of 8-35 K in the cooling process, but it has dramatic quenching in 8-35 K in heating process and is followed by antiquenching in 35-225 K and quenching in 225-290 K. Therefore sample $S_{4\%}$ shows the most prominent and complex transition behavior between quenching and antiquenching as well as the remarkable PL hysteresis during cooling and heating processes; returning back to room temperature half of the PL intensities for both emissions is lost. For $S_{8\%}$, blue emission shows quenching in 290-250 K and weak antiquenching in 8-250 K in cooling process while it is quenching in heating process; the orange emission is antiquenching in 50-225 K in cooling process. The PL hysteresis for this sample is also very strong and intensity drop at room temperature is even more.

Understanding of the mechanism of antiquenching and hysteresis for the ZnS/CdS:Mn/ZnS QWQDs needs the knowledge of energy transfer processes and radiative nature of the emissions. The well prepared CdS:Mn/ZnS QDs demonstrate a very sharp blue emission peak at 420 nm which is assigned to the emission from exciton recombination in QDs.[32,11] While trapping of delocalized charge carriers in defect states is considered efficient and competes with the radiative recombination of excitons; then the defect-related emission should be stronger than exciton emission in QDs, which shows the a broad emission band centered at 430 nm.[14] However, contradictory results on the temperature dependence of CdS:Mn QDs exist in the literatures: both the blue and orange emissions were found thermal quenching;[14] but another study has shown an indication of antiquenching for orange emission and quenching for blue emission (Fig. 6 in ref. [13]), which resembles to the present case of $S_{6\%}$ (Fig. 3(b)). Therefore, an antiquenching behavior should be only indirectly related to the structure and $Mn^{2+}$ doping level but more directly to the nature of defects which may vary with sample preparation method and condition. The defect-related blue emission in ZnS is considered to be a radiative recombination of either a delocalized charge carrier (electron) with a trapped charge carrier (hole)[13] or a trapped electron (donor) with a trapped hole (acceptor)[14] for donor-acceptor emission.[33] Trapping of the holes occurs at sulfur vacancies and subsequent recombination with electrons gives to the blue emission. As $Mn^{2+}$ ions are hole attractive, holes can also be trapped to $Mn^{2+}$ and the recombination with electrons enhances the orange $Mn^{2+}$ emission by providing additional channel of energy transfer from ZnS host to $Mn^{2+}$.[13] Because of the large lattice mismatch for the ZnS/CdS/ZnS QWQDs,[23] considerable defects should exist in the ZnS/CdS interface where the strongly bound $Mn^{2+}$ is also likely rich. This explains why the blue emission intensity here seems also to relate to $Mn^{2+}$ concentration at room temperature (Fig. 3) but for ZnS:Mn quantum rods, for which the blue emission was considered from surface defects, the intensity only weakly changes with $Mn^{2+}$ doping level.[16]

The antiquenching for orange emission from multishell CdSe/CdS/ZnS QDs was explained according to the enhanced emission from release of localized charge carriers by thermal activation assuming that they are firstly trapped at the core-shell and shell-shell interfaces after excitation due to energy band fluctuation induced by the lattice strain at the interfaces.[30] In view of this, the full temperature dependence is described by an equation,

$$I(T) = I_0 \frac{1 + C\exp(-\Delta E/k_B T)}{1 + A\exp(-E_a/k_B T) + B\left[\exp(E_{LO}/k_B T) - 1\right]^{-m}} \tag{1}$$

where $E_a$, $E_{LO}$ and $\Delta E$ represent, respectively, the activation energy of defects for nonradiative relaxation, LO phonon energy and energy depth of the localized charge traping states; $m$ is number of LO phonons for assisted thermal escape of carriers from QWQDs. The $A$ and $B$ terms in the denominator represent the normal quenching behavior, while the $C$ term is due to the extra emission from activation of antiquenching centers and accounts for the antiquenching behavior. Taking $E_{LO} = 43$ meV for ZnS,[34] Eq. (1) with one or two $A$-terms can well describe the temperature dependence for both blue emission and $Mn^{2+}$ orange emission and in both cooling and heating processes, as shown in Fig. 3. The charge trap states fitted have shallow depths (typically $\Delta E$ is about 10-30 meV) and can be thermally activated. Having an additional energy transfer channel to $Mn^{2+}$ (Eqs. (3)-(6) in ref. [35]), the blue emission is under different carrier relaxation dynamics with $Mn^{2+}$ orange emission and therefore should be with different parameter sets for quenching and antiquenching; then a competition between two emissions exist, which results in their interrelated quenching and antiquenching behavior. For example, for $Mn^{2+}$ emission from $S_{4\%}$ in heating process, the competition between energies of $E_a = 3$ meV and $E_{LO} = 43$ meV for quenching and energy of $\Delta E = 16.3$ meV for antiquenching displays the transition from quenching to antiquenching at ~50 K (4.3 meV) and transition from antiquenching to quenching at 220 K (19 meV).

The hysteresis phenomenon is considered to be specifically from the defects in the QWQDs. Under external hydrostatic pressure, PL of core-sell QDs is usually blue shifted [37,38]. However, experiment and theoretical calculation indicates it is redshift for $Mn^{2+}$ emission from ZnS:Mn and CdS:Mn/ZnS QDs.[36,39] Owing to 7% lattice mismatch between CdS core and ZnS shell, the pressure in CdS/ZnS QDs is estimated to be more than 4 GPa for 7.5 monolayers of ZnS.[36] Here, $S_{4\%}$, $S_{64\%}$ and $S_{8\%}$ have roughly a linear redshift for $Mn^{2+}$ peak down to 50 K, which is completely reversible after returning to room temperature. The largest redshift of -3.5 meV corresponds to an increase of internal pressure by ~0.1 GPa, estimated with pressure rate of ~-33.3 meV/GPa,[39] at low temperature due to enhanced lattice contraction by temperature besides that by lattice mismatch and surface tension. This could possibly introduce more defect states at ZnS/CdS interfaces where $Mn^{2+}$ is considered rich to create more quenching centers and/or to lower their activation energy. In heating process, PL intensity can thus be more quickly quenched to present hysteretic PL temperature dependence.

In summary, large antiquenching effect has been observed for blue emission and $Mn^{2+}$ orange emission from ZnS/CdS:Mn/ZnS QWQDs. The effect is explained as an extra emission from release of localized charge carriers by thermal activation, and can be well described by an equation incorporating usual quenching terms and an antiquenching term. Competitions between quenching and antiquenching mechanisms and between two emissions due to different energy transfer channels, together results in their complex and intercorrelated PL temperature dependences. The irreversible PL intensity during cooling and heating processes may be related to the hysteretic creation of quenching centers and/or reduction of their activation energy induced at the ZnS/CdS core-shell by the lattice contraction at low temperature for the large lattice mismatched QWQDs.

This work was supported by the National Natural Science Foundation of China (Grant Nos. 11074232, 11274288 and 21002097), the National Basic Research Program of China (Nos. 2011CB932801 and 2012CB933702) and Ministry of Education of China (No. 20123402110034).

**Figure Captions**

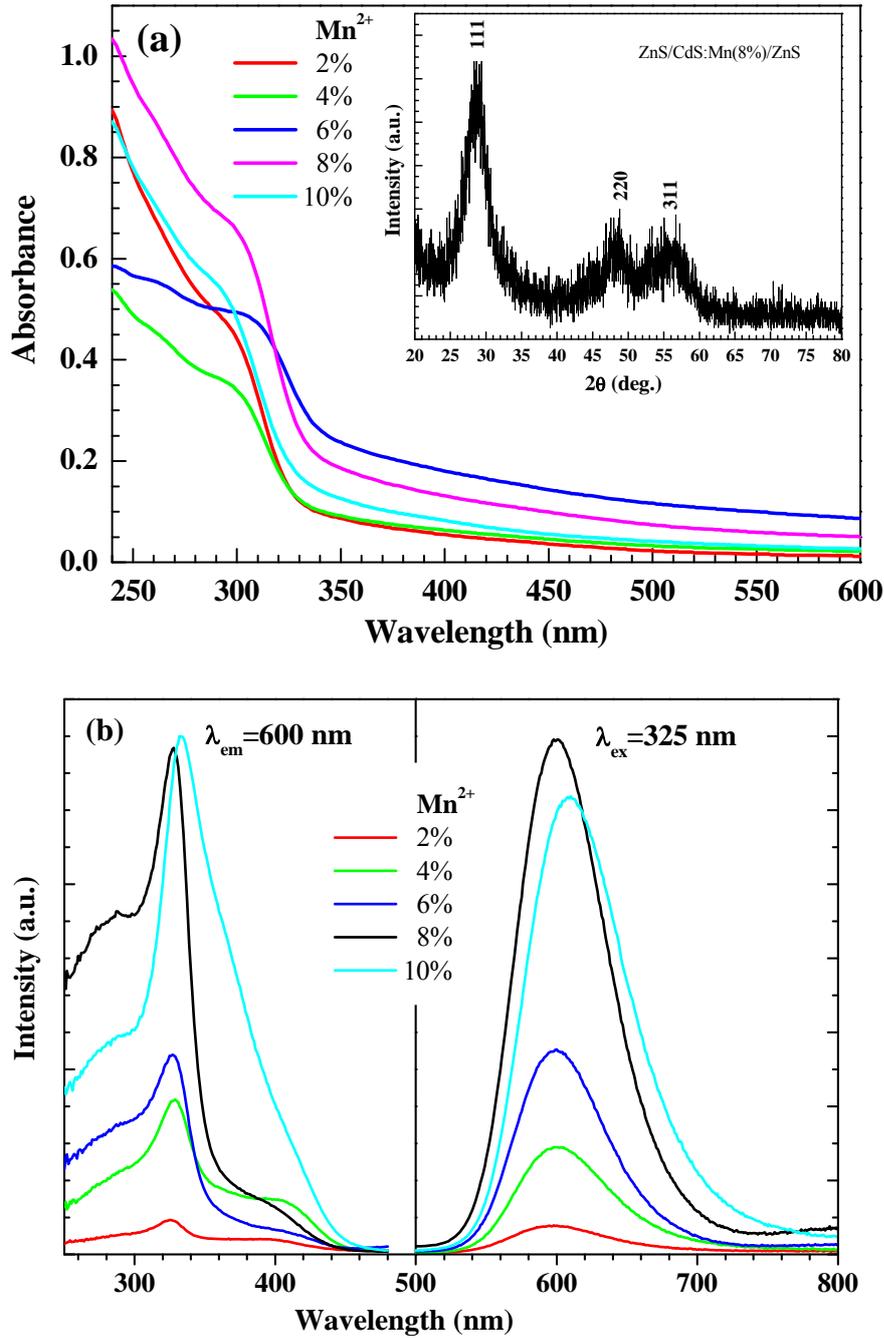

Fig. 1 (a) UV-Vis absorbance spectra and (b) PLE (left) and PL (right) spectra of as-grown ZnS/CdS:Mn/ZnS QWQDs for different initial $Mn^{2+}$ concentrations. The inset in (a) shows x-ray diffraction pattern for $S_{8\%}$.

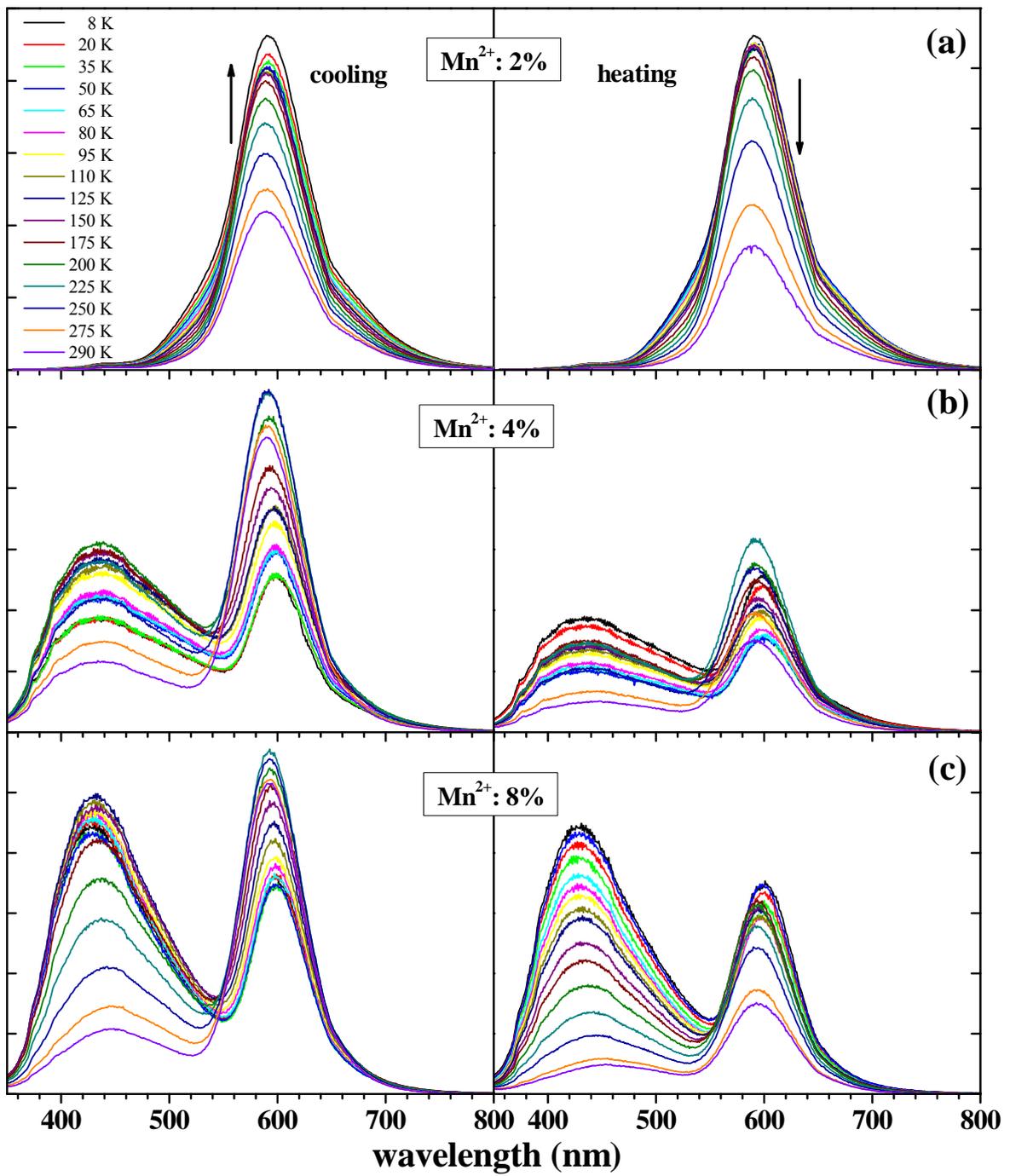

Fig. 2 Temperature dependent PL spectra of ZnS/CdS:Mn/ZnS QWQDs for samples: (a) $S_{2\%}$; (b) $S_{4\%}$; (c) $S_{8\%}$. Left panel is for temperature cooling process and the right panel for temperature heating process in the range of 8-290 K.

**Fig. 3**

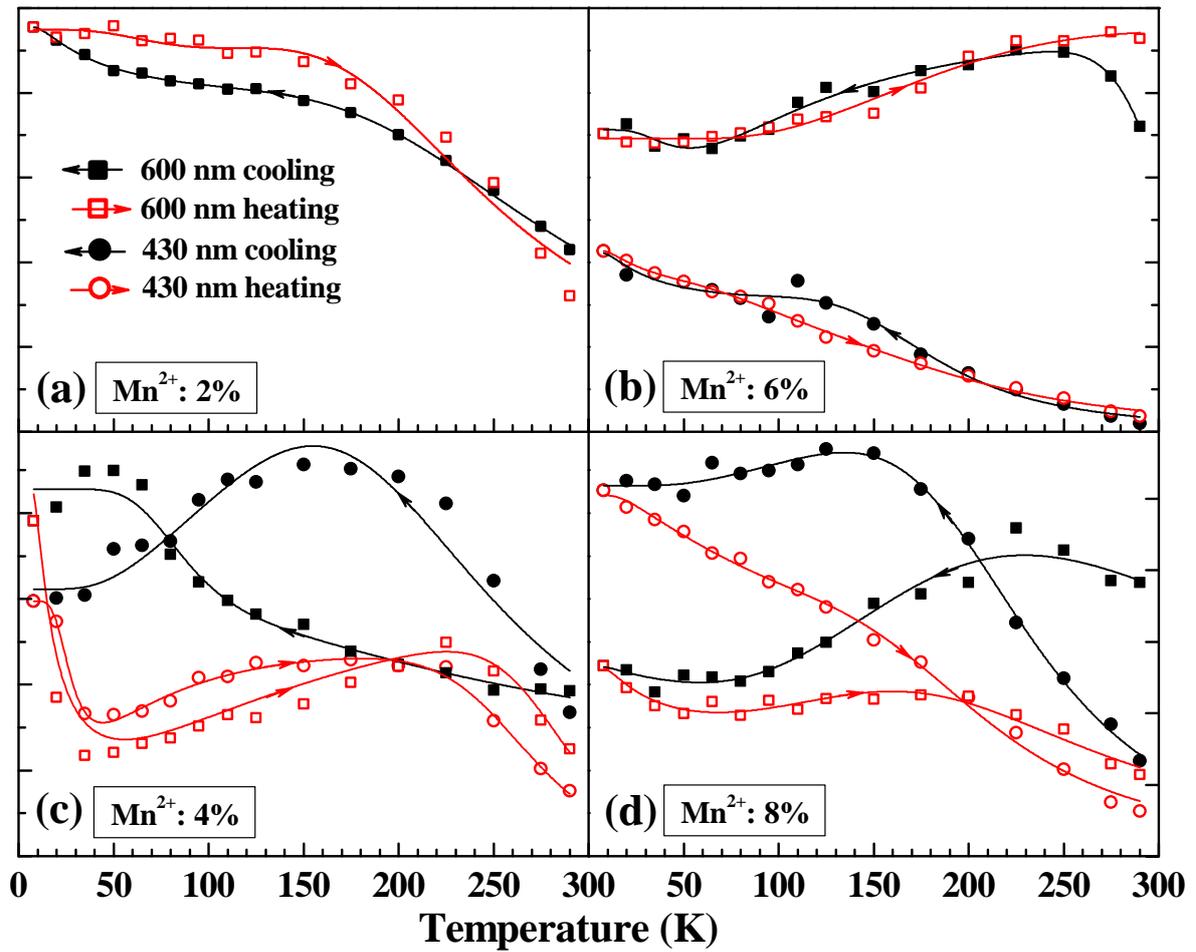

Fig. 3  Temperature dependence of integrated PL intensity for blue emission of 430 nm (circle) and orange emission of 600 nm (square) of ZnS/CdS:Mn/ZnS QWQDs for samples: (a) $S_{2\%}$; (b) $S_{6\%}$; (c) $S_{4\%}$; (d) $S_{8\%}$. Close symbols represent for the temperature cooling process and open symbols for heating process. Curves are fitting to the experimental data by Eq. (1).